\documentclass[manuscript]{acmart}

\usepackage{hyphenat}
\DeclareUnicodeCharacter{FE0F}{}
\DeclareUnicodeCharacter{203C}{!!}
\usepackage[utf8]{inputenc}
\usepackage{newunicodechar}
\usepackage{hyperref}
\usepackage{booktabs}
\usepackage{multirow}
\usepackage{wrapfig}
\usepackage{tabularx}
\usepackage{subcaption}
\usepackage{xspace}
\usepackage{placeins}
\usepackage[normalem]{ulem}
\newcommand{\corpus}{\textsc{TwistedHumor}\xspace}

\newif\ifshowcomments
\showcommentsfalse% change to \showcommentstrue to show comments

\AtBeginDocument{%
  }

\begin{document}

\title{When Jokes Cross the Line: Analyzing Regular Humor and Dark Humor in YouTube Shorts}
%No dataset in title 

\author{Sydney Danielle Johns}
\email{sydneyjohns@vt.edu}
\authornotemark[1]
\affiliation{%
  \institution{Virginia Polytechnic Institute and State University}
  \city{Alexandria}
  \state{Virginia}
  \country{USA}
}

\author{Sanjeev Parthasarathy}
\email{sanjeev26@vt.edu}
\affiliation{%
  \institution{Virginia Polytechnic Institute and State University}
  \city{Alexandria}
  \state{Virginia}
  \country{USA}
}

\author{Shantnu Bhalla}
\email{shantnub@vt.edu}
\affiliation{%
  \institution{Virginia Polytechnic Institute and State University}
  \city{Alexandria}
  \state{Virginia}
  \country{USA}
}

\author{Vaibhav Garg}
\email{vaibhavg@vt.edu}
\affiliation{%
  \institution{Virginia Polytechnic Institute and State University}
  \city{Alexandria}
  \state{Virginia}
  \country{USA}
}

%%
%% By default, the full list of authors will be used in the page
%% headers. Often, this list is too long, and will overlap
%% other information printed in the page headers. This command allows
%% the author to define a more concise list
%% of authors' names for this purpose.
%\renewcommand{\shortauthors}{Trovato et al.}

%%
%% The abstract is a short summary of the work to be presented in the
%% article.

\begin{abstract}
Video platforms such as YouTube have reshaped how users engage with entertainment and information, emphasizing brief, highly engaging content such as Shorts. Within this ecosystem, certain content occupies a gray area where it remains allowed but may still have unintended negative effects on some audiences. To study this problem, we introduce \corpus, a dataset of 1,211 YouTube Shorts paired with 33,041 related comments, with hand annotations for humor presence, humor type, harm, topic, rhetorical devices, and stand up context. Beyond dataset creation, we present a multi view analysis of how humor and harm appear in short form social media. Using LLooM based concept induction over video descriptions, we find that dark humor frequently clusters around themes of critique, coping, awkwardness, and identity expression rather than appearing as a single uniform category. We further analyze audience response through linked comments and show that regular humor is associated with more positive sentiment, while dark humor receives more mixed, neutral, and sometimes more toxic reactions. Finally, we evaluate large language models against human annotations and find that they perform better on stand up comedy compared to shorter jokes. Together, these results position \corpus not only as a new benchmark, but as an empirical study of the gray area between humor and harm in short form video, highlighting the need for context aware moderation and more robust multimodal evaluation.
\end{abstract}

%\keywords{YouTube Shorts, Dark Humor, Harmful Content Detection, Multi-modal Dataset, Social Media Moderation, Inter Annotator Agreement, Large Language Models, Sentiment Analysis, Toxicity Detection}

%%
%% This command processes the author and affiliation and title
%% information and builds the first part of the formatted document.
\begin{CCSXML}
<ccs2012>
   <concept>
       <concept_id>10003120</concept_id>
       <concept_desc>Human-centered computing</concept_desc>
       <concept_significance>500</concept_significance>
       </concept>
   <concept>
       <concept_id>10010405</concept_id>
       <concept_desc>Applied computing</concept_desc>
       <concept_significance>500</concept_significance>
       </concept>
 </ccs2012>
\end{CCSXML}

\ccsdesc[500]{Human-centered computing}
\ccsdesc[500]{Applied computing}

\maketitle 

\section{Introduction}

The rise of short-form videos has reshaped how users consume entertainment and, increasingly, information, with platforms such as YouTube, Instagram, and TikTok. Instagram and TikTok reels, and YouTube Shorts, provide highly engaging content \cite{molem2024keepin, zannettou2024tiktok, ling2022virality}. Among these platforms, YouTube remains one of the most widely used, with over 80\% of adults in the United States reporting usage \cite{park_americans_2025}. The platform encompasses many aspects of everyday life, including teaching new skills, providing updates on current events, and serving as a primary source of entertainment. YouTube also hosts a large volume of short form content through YouTube Shorts, which are limited to 180 seconds \cite{kumar_latest_2025}. Given the volume of entertainment available on YouTube, many users report ``doomscrolling,'' or watching content for longer than they intended \cite{violot_shorts_2024}. 

Prior studies argue both users and moderators should consider the safety and the broader impact of the content being shared on these platforms \cite{jhaver2023personalizing, moscato2025personalization}. YouTube guidelines are effective at identifying clearly harmful or policy-violating material \cite{youtube_guidelines}. However, a substantial portion of content persists in a gray area where it is permitted but may still have unintended negative effects on certain audiences \cite{milton2023iseeme, chu2026bigtokdetect, nguyen2025supporters}. This limitation is particularly evident in cases such as dark humor, where content may be humorous and permissible, yet still potentially disturbing, especially to the community being targeted \cite{ali2024kms, allison2019humour}. 

Dark humor is a widely consumed form of entertainment that often engages with taboo or sensitive subjects \cite{kasu_d-humor_2025}. Prior research shows that dark humor targets multiple communities based on attributes such as \textit{religion, race, sex, nationality, and disability} \cite{zia2021memes}. While some viewers perceive such content as humorous, others may find it distressing or triggering, depending on individual experiences and sensitivities \cite{niu2026behindmeme, chen2025suicidememes}. This motivates the need for mechanisms that alert users in advance about such potentially sensitive content \cite{yang2025behavioral}. 

% Importantly, short-form videos typically provide little to no prior context or signaling about the nature of such content. As a result, users may encounter potentially sensitive material without warning, particularly in high-volume, rapid-consumption settings such as doomscrolling. 

% This variation in audience response reflects a deeper challenge in distinguishing humor from harm in such content\vg{this we already covered in second para}. Unlike explicitly harmful or hateful content, which is defined by clear violations such as targeting protected attributes, harmful content in these contexts often depends on interpretation, intent, delivery, and audience perception \cite{martinez_pandiani_toxic_2025}. As a result, dark humor should not be assumed to be inherently harmful. Instead, we treat harm as a separate analytical dimension that captures when a joke’s framing, target, or social meaning introduces the potential for negative impact. This framing allows us to move beyond binary notions of “allowed vs. disallowed” content and instead study how humor and harm interact in ambiguous, real-world settings.

These challenges highlight the need to better understand how humor is expressed, interpreted, and received in short-form social media content. To address this, we investigate the following research questions:
\begin{itemize}

\item \textbf{RQ1:} How do topics, target categories, and rhetorical devices differ between regular humor and dark humor in short-form videos?

To address RQ1, we construct a dataset of 1,211 YouTube Shorts with detailed human annotations, including \emph{joke topic}, \emph{target category}, and \emph{rhetorical device}. We analyze these attributes to characterize how dark humor differs from regular humor in terms of subject matter, targeted groups, and stylistic framing.

\item \textbf{RQ2:} How accurately can existing LLMs identify humor presence and distinguish regular humor from dark humor using video transcripts?

To address RQ2, we evaluate multiple large language models (LLMs), including GPT-5 Mini, Claude Sonnet 4.6, Gemini 2.5 Flash, Gemini 3.1 Pro Preview, and DeepSeek V3.1, on transcript-based humor understanding tasks. Specifically, we measure model agreement with human annotations across \emph{humor presence} and \emph{humor type}, enabling us to assess how well current models capture the distinction between regular and dark humor.

\item \textbf{RQ3:} How do audience responses, measured through comment sentiment, emotion, and toxicity, vary between regular humor and dark humor videos?

To address RQ3, we analyze audience responses using the top 30 comments associated with each video. We compute sentiment, emotion distributions, and toxicity scores, and examine how these signals differ across humor types. We further use LLooM-based concept induction to explore higher-level patterns in humor themes and audience interpretation.

\end{itemize}

\section{Related Work}

Prior work on short form and multimodal media provides useful foundations, but does not directly address the distinction between humor, dark humor, and harm. TikTokActions studies TikTok clips for human action recognition rather than contextual interpretation \cite{qian_advancing_2024}, while other work examines engagement signals such as likes, views, shares, and comments to study viewer response \cite{ziyada_video_2024}. Similarly, MultiVENT provides multilingual videos of real world events with aligned natural text, but its focus is event understanding and cross lingual retrieval rather than humor \cite{sanders2023multiventmultilingualvideosevents}. Humor datasets have also largely come from structured settings rather than social media feeds. \textit{UR-FUNNY} uses TED talks for multimodal humor understanding \cite{hasan_ur-funny_2019}, \textit{When to Laugh and How Hard} uses the \textit{Friends} TV show for humor detection and intensity estimation \cite{alnajjar2022when}, and MUStARD studies multimodal sarcasm in television dialogue \cite{castro_towards_2019}, with later work extending it to emotion recognition in sarcasm \cite{ray_multimodal_2022}. . These datasets support humor and sarcasm modeling, but they do not capture the challenges of short form social media.

Recent work has begun to study humor understanding in user generated short form videos. ExFunTube examines humorous moments in YouTube short form videos and explores prompting strategies for humor understanding \cite{ko_can_2024}, while YouNiverse provides large scale YouTube metadata useful for analyzing platform level engagement \cite{ribeiro_youniverse_2021}. Prior work has also analyzed YouTube comments on stand up comedy videos using an LSTM model \cite{supriyono_analyzing_2024}. These studies motivate our use of transcript based evaluation, metadata, and comment analysis, but they do not focus on separating regular humor, dark humor, and harmful content in the same framework.

Research on harmful media highlights the difficulty of separating harm from hate in ambiguous cases. \textit{Detecting Harmful Memes and Their Targets} argues that harm can require contextual judgment beyond keyword matching and emphasizes the importance of identifying who or what is targeted \cite{pramanick_detecting_2021}. Dark humor adds further complexity because it often depends on sensitive, implicit, and culturally specific cues. D-HUMOR treats dark humor as distinct from sarcasm and introduces a multimodal dataset with labels for dark humor presence, target category, and intensity \cite{kasu_d-humor_2025}. These works motivate our focus on short form videos.

Recent work on humor in the media has mainly focused on humor detection tasks, such as identifying whether content is humorous or not \cite{castro_towards_2019,tong_hummus_2026, liu2024funnynetwmultimodallearningfunny, Rehman_2025}. These studies do not fully examine the boundary case between dark humor and genuinely harmful content, where meaning depends on context, rhetoric, delivery, and audience interpretation. This limitation is especially important in short form social media, where brief clips can blend irony, satire, shock, and harm in ways that are difficult to separate with simple humor versus non humor labels. Table~\ref{tab:dataset_size_comparison} shows key differences between \corpus and similar work. Our work not only introduces a hand annotated dataset of YouTube Shorts but also analyzes humor and audience response. The contribution is more than just a new dataset, we complete an analysis of the gray area between dark humor and harm in short form video.

\begin{table*}[t]
\centering
\caption{Comparison with related multimodal and social media datasets. Hand-annotated video benchmarks are often in the low-thousands rather than at the scale of large text corpora.}
\label{tab:dataset_size_comparison}
\small
\begin{tabular*}{\textwidth}{@{\extracolsep{\fill}} p{2.4cm} p{2.8cm} p{2.4cm} c p{5.2cm}}
\toprule
\textbf{Dataset} & \textbf{Modality} & \textbf{Source} & \textbf{Size} & \textbf{Task / Focus} \\
\midrule
MUStARD \cite{castro_towards_2019} 
& Video, audio, text 
& TV shows 
& 690 
& Multimodal sarcasm detection \\

HateMM \cite{das_hatemm_2023} 
& Video, audio, text 
& BitChute, Odysee 
& 1,083 
& Hate vs.\ non-hate video classification \\

Hummus \cite{tong_hummus_2026} 
& Image, text 
& Web / social media style content 
& 1,000 
& Multimodal humor understanding \\

\textbf{\corpus} 
& Video, audio, transcript, metadata, comments 
& YouTube Shorts 
& \textbf{1,211} 
& Humor, dark humor, and harm in short-form social media \\

MultiHateClip \cite{wang_multihateclip_2024} 
& Video, audio, text 
& YouTube, Bilibili 
& 2,000 
& Hateful, offensive, and normal short-video classification \\

\bottomrule
\end{tabular*}
\end{table*}

\section{Dataset Collection}

We developed a pipeline to collect YouTube Shorts without using the official YouTube API. First, we compiled curated keyword lists and used a Python script to generate search URLs directly from those terms by identifying patterns in YouTube’s URL structure. A key engineering contribution of this pipeline was identifying the YouTube search filter codes needed to target Shorts content. In total, we generated 220 YouTube search URLs from these curated keyword lists. Next, we loaded each YouTube search results page and saved the raw HTML source code. We then performed structured parsing to extract the video IDs, titles, durations, and view counts. To ensure sufficient audience exposure, we retained only videos with at least 10,000 views. Overall, the automated pipeline collected 1,997 candidate videos, from which we sampled a final set of 1,211 unique videos for annotation and analysis.

\subsection{Keyword Based Retrieval Strategy}
To discover candidate videos, we used a creator and channel based queries consisting of widely recognized comedians and humor focused channels. These creators were identified using publicly available rankings based on subscriber counts \cite{top_10_YT,top_20_standup,top_30_comedy}. Creator and channel queries help retrieve candidate Shorts from accounts that consistently publish comedic material. In addition to creator based retrieval, we included topical keyword queries organized into four broad subject groups: Health and Safety, Politics and Society, Conflict and Global Events, and Environmental and Ethical Issues. These topical keywords were selected to find socially relevant videos. Keyword based retrieval has been used in prior dataset construction to gather candidate samples before applying manual annotation. For example, the GoEmotions dataset used keyword queries to collect Reddit comments likely to contain emotional language, which were then manually labeled \cite{demszky_goemotions_2020}. Similarly, HateXplain collected candidate social media posts containing potentially harmful language and relied on human annotators to determine the final labels \cite{mathew_hatexplain_2022}. Keywords serve only as a discovery mechanism for candidate videos. Final dataset labels and content categorization were determined through manual annotation rather than keyword matching. Using multiple query groups reduces the risk that the dataset reflects a single topic distribution or creator community. 

\begin{comment}
Keyword lists are provided in Table \ref{tab:creator_channel_keywords}.

\begin{table}[t]
\centering
\caption{Creator and Channel Based keywords}
\label{tab:creator_channel_keywords}
\begin{tabular}{p{15cm}}
\toprule
\textbf{Creator-Based Keywords} \\
\midrule
Ellen DeGeneres, Kevin Hart, Joe Rogan, Matt Rife, Steve Harvey, Nigel Ng, Trevor Noah, Russell Brand, Adam Sandler, Theo Von, Gabriel Iglesias, Jamie Foxx, Trevor Wallace, Ricky Gervais, Amy Schumer, Andrew Schulz, Bill Maher, Chelsea Handler, Sarah Silverman, Chris Rock \\
\midrule
\textbf{Channel-Based Keywords} \\
\midrule
Smosh, David Dobrik, College Humor, First We Feast, Funny or Die, The Tonight Show Starring Jimmy Fallon, PewDiePie, The Late Night Show, Good Mythical Morning, Tana Mongeau, hhProductions, Dolan Twins, JennaMarbles, Emma Chamberlain, Markiplier, Good Mythical MORE, Jacksfilms, Nigahiga, Liza Koshy, Shane Dawson TV, The Try Guys, Gus Johnson, Cody Ko, Danny Gonzalez, Drew Gooden, Kurtis Conner, Penguinz, Ryan Trahan, MrBeast, CalebCity, LongBeachGriffy, King Bach, NELK, Trevor Wallace, Mr Beast Gaming, RDCworld, ImDontai, Internet Historian, TheOddsOut, Jaiden Animations, SMG, Brent Rivera, Zach King, Jordan Matter, Lele Pons, Kevin Langue, Josh Joshson, Don't Tell Comedy, Netflix Is A Joke, Comedy Central Stand-Up, Jimmy O Yang \\
\bottomrule
\end{tabular}
\end{table}

\end{comment}

\subsection{Dataset Assembly}
Each dataset record links a video to its extracted platform metadata, transcript text, comments, and description. The final dataset is released in CSV format, where each row corresponds to a single YouTube Short. This CSV combines manual annotations with metadata collected during retrieval. To preserve the original text data, comments and descriptions are also stored separately in folders organized by video ID. Annotation fields capture the humor related properties of each video. These include \texttt{humor\_presence}, which indicates whether humor is present in the video, and \texttt{humor\_type}, which distinguishes between regular humor and dark humor. The field \texttt{joke\_topic} identifies the primary topics referenced in the joke, while \texttt{rhetorical\_device} records the use of rhetorical mechanisms such as irony or satire. The field \texttt{stand\_up} indicates whether the video contains stand up comedy. For videos involving potentially sensitive targets, the field \texttt{target\_category} records the entity or group referenced by the humor. For each video, we used the Whisper model \cite{Whisper} to transcribe the audio as \texttt{transcript\_text} along with the video identifier \texttt{video\_id} and direct video link \texttt{url}. Additional metadata collected from YouTube includes the video \texttt{title}, \texttt{tags}, as well as channel information such as \texttt{channel}, \texttt{channel\_id}, \texttt{uploader}, and \texttt{uploader\_id}. Temporal metadata includes the \texttt{upload\_date} and video \texttt{duration}. Engagement statistics such as \texttt{view\_count}, \texttt{like\_count}, and \texttt{comment\_count} are also included. Finally, each video is detected \texttt{language}, and the original \texttt{searched\_keyword} used during data collection.

%ADD A PICTURE 

\section{Annotation Process}
\subsection{Annotation Platform and Agreement}
Annotations were completed in Label Studio \cite{LabelStudio} using a project standard of procedure (SOP) document as a guide. We deployed the Label Studio annotation platform within an AWS Virtual Private Cloud (VPC) to provide a controlled and secure environment for the labeling process, while the dataset videos themselves were stored in Amazon S3 for reliable and scalable access during annotation. There are also fixed label definitions for regular and dark humor. Regular humor refers to multimodal content that creates amusement through dichotomy, surprise, or playful reinterpretation of meaning. It may employ rhetorical devices such as irony or satire but these devices operate within socially acceptable, non-sensitive topics that most people can collectively laugh at. \cite{kasu_d-humor_2025} Dark humor is multimodal humor that uses irony, satire, sarcasm, or cynicism to produce amusement through taboo, offensive, or culturally sensitive themes. It relies on implicit, context-dependent cues, visual–textual dichotomy, and emotionally conflicting or morally provocative elements. \cite{kasu_d-humor_2025}. Annotators completed training and were required to learn and understand the formal definitions and discuss edge cases, especially the boundary between regular humor and dark humor. After training, annotators completed three rounds of discussions to settle any disagreements. The three annotators are authors of this paper. 

First, three annotators, independently labeled an initial set of 200 videos. After achieving satisfactory agreement on this set, the next 200 videos were labeled by two annotators. Once agreement stabilized on the first 400 videos, the remaining 811 videos were labeled by a single annotator. To quantify labeling consistency, we computed inter annotator agreement (IAA) on phases of the dataset. For the initial 200 video set labeled by three annotators, we report Krippendorff's alpha, which supports agreement estimation for more than two annotators and does not assume equal class prevalence \cite{krippendorff_1970}. For the subsequent 200 videos labeled by two annotators, we report Cohen's kappa, which measures agreement beyond chance for paired ratings \cite{cohen_1960}. Humor presence achieved moderate agreement ($\alpha = 0.61$), indicating reasonable consistency in determining whether a video contained humor. Humor type showed stronger agreement ($\alpha = 0.72$), suggesting that annotators were generally consistent in distinguishing these categories once humor was identified. Stand-up classification achieved near-perfect agreement ($\alpha = 0.96$), indicating that this label was clearly defined and consistently applied. Target category labeling also showed strong agreement ($\alpha = 0.83$). Overall, these results suggest that structural labels such as stand-up and target category can be annotated reliably.

\subsection{Annotation Workflow}
The annotation workflow follows these steps:
\begin{enumerate}
    \item Humor presence (Humor vs Not Humor). If Not Humor, annotators skip the remaining steps.
    \item Joke topic selection (at least one topic, or Other).
    \item Rhetorical device selection (Irony, Satire, or Neither).
    \item Humor type (Regular Humor vs Dark Humor).
If Dark Humor is selected, annotators additionally label:
    \begin{itemize}
        \item Target Category (Gender or Sex related, Mental Health, Disability, Race or Ethnicity, Violence or Death, Other Sensitive Target)
        \item Intensity on a three level  (Mild, Moderate, Severe)
    \end{itemize}
\end{enumerate}

\section{Dataset Statistics}
The final dataset consists of 1{,}211 unique YouTube Shorts collected from 716 distinct channels. Transcripts vary in length, with an average of 140.65 words and a median of 126 words. Video titles are relatively short, averaging 8.13 words, while descriptions are longer, with an average of 34.24 words. The dataset has a mix of humorous and non-humorous content. Among the 1,211 videos, 601 (49.62\%) were annotated as humorous and 610 (50.37\%) as non-humorous or ambiguous. When examining humor type, 402 videos (33.20\%) were labeled as regular humor and 199 (16.43\%) as dark humor. 

Stand-up comedy represents a smaller portion of the dataset, with 176 videos (14.53\%) labeled as stand-up and 1{,}032 (85.22\%) as non-stand-up. Among stand-up videos, 96 were labeled as dark humor, representing 54.55\% of all stand-up posts. Regular humor accounted for 78 stand-up posts. This shows that dark humor was slightly more common than regular humor within the stand-up subset. 

\begin{figure}[t]
    \centering

    % --- (a) Joke Topic ---
    \begin{subfigure}[b]{0.52\linewidth}
        \centering
        \includegraphics[width=\linewidth]{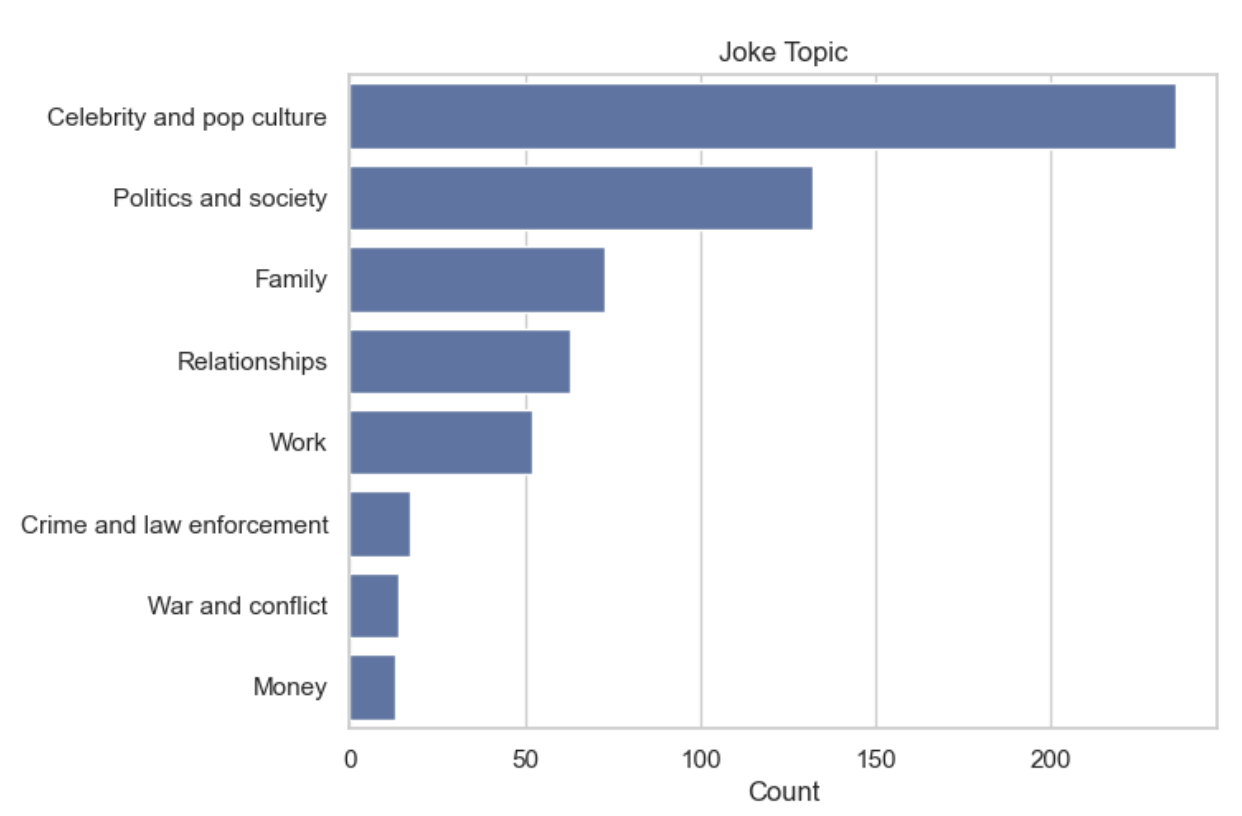}
        \caption{Joke topics.}
        \label{fig:joke_topic}
    \end{subfigure}
    \hfill
    % --- (b) Description Word Count ---
    \begin{subfigure}[b]{0.44\linewidth}
        \centering
        \includegraphics[width=\linewidth]{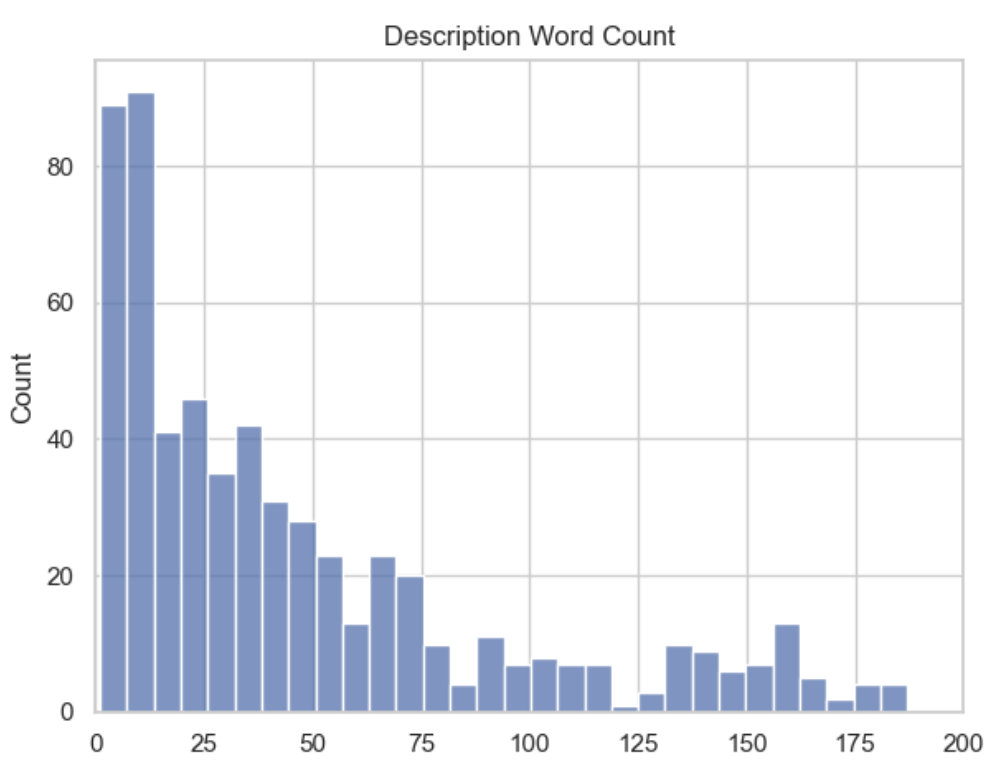}
        \caption{Description word count.}
        \label{fig:word_count}
    \end{subfigure}

    \caption{Distributions of selected dataset variables in TwistedHumor. Panel (a) shows the frequency of the most common joke topics. Panel (b) shows the distribution of description word counts, indicating that most videos have short descriptions.}
    \Description{Distributions of selected dataset variables in TwistedHumor. Panel (a) shows the frequency of the most common joke topics. Panel (b) shows the distribution of description word counts, indicating that most videos have short descriptions.}
    \label{fig:key_dataset_variables}
    \vspace{-0.5cm}
\end{figure}

Most videos do not target a specific sensitive group, with 86.29\% of entries having no annotated target category. Among videos that do include a target, the most common categories are race or ethnicity (4.46\%), gender or sex-related topics (2.64\%), and other sensitive targets (2.06\%). Videos typically contain a small number of topics, with an average of 1 topic per video. The most common topics include Celebrity and pop culture, and Politics and society. Figure~\ref{fig:key_dataset_variables} shows the distribution of joke topics and description length. Overall, these statistics demonstrate that \textsc{} contains a diverse range of humor topics and varied video descriptions, supporting its use for studying humor across different content themes and levels of contextual detail.

% \begin{figure}
%     \centering
%     \includegraphics[width=\linewidth]{output6.png}
%     \caption{Distributions of key dataset variables in TwistedHumor. The top left panel shows the distribution of humor presence labels, with the dataset containing slightly more videos labeled as not humor than humor. The top right panel shows the frequency of the most common joke topics, where celebrity and pop culture is the dominant topic. The bottom left panel shows the stand up distribution, indicating that only a minority of videos are stand up clips. The bottom right panel shows the distribution of description word counts which shows that most videos have short descriptions. Together, these panels highlight the diversity of the dataset.}
%     \Description{A four panel figure summarizing key variables in the TwistedHumor dataset.}
%     \label{fig:key_dataset_variables}
% \end{figure}

Hashtag usage further illustrates the strong comedic orientation of the dataset. The most frequently occurring hashtags were \#shorts  with 340 mentions, \#funny with 260 mentions, and \#comedy with 134 mentions. Stand up related tags such as \#standupcomedy and \#standup also appeared frequently. General visibility and platform engagement tags such as \#trending, \#viral, \#fyp, and \#youtubeshorts were also common. A smaller set of hashtags referenced political content, such as \#trump, \#usa, and \#news. The hashtag distribution supports the dataset's mix of humor focused, creator centered, and socially relevant content.

\section{Analysis}
\subsection{Evaluating Humor Understanding in Large Language Models}
We evaluated agreement between LLM predictions and human annotations across \emph{humor presence}, \emph{humor type}, \emph{stand-up classification}, and \emph{target category}. Table~\ref{tab:llm_full_dataset} reports performance for each model. Across models, humor presence classification remained challenging. Accuracy ranged from $0.50$ to $0.58$, indicating that models often struggled to reliably determine whether humor was present in short-form videos. Gemini 3 Flash Preview achieved the highest humor presence accuracy ($0.576$), followed by DeepSeek V3.1 ($0.564$), GPT-5 Mini ($0.542$), Gemini 2.5 Flash ($0.534$), and Claude Sonnet 4.6 ($0.501$). These results suggest that even strong language models have difficulty identifying humor when visual delivery, timing, and context are absent.
% \vspace{-10pt}

Humor type classification was similarly difficult. Accuracy ranged from $0.458$ to $0.502$. The narrow spread across models suggests that distinguishing regular humor from dark humor remains a difficult task even when humor is already detected. This result aligns with the annotation analysis discussed earlier, where even human annotators show lower agreement on this task. In contrast, stand-up classification produced the strongest performance across models, with accuracies ranging from $0.681$ to $0.850$. Gemini 3 Flash Preview achieved the highest stand-up accuracy followed closely by Claude Sonnet 4.6. This pattern suggests that stand-up comedy contains recognizable structural and lexical cues. Target category prediction was also comparatively stronger, with accuracies ranging from $0.746$ to $0.807$. Taken together, these results highlight that models perform best on structural or more explicitly signaled attributes, such as stand-up format, but struggle with higher-level interpretive tasks such as humor detection. In particular, distinguishing dark humor from regular humor remains difficult even for advanced models. This finding reinforces the motivation for our dataset, which focuses on ambiguous cases where humor, harm, and interpretation intersect. 

\FloatBarrier
\begin{table}[!h]
\centering
\caption{Transcript-only model performance on the full 1,211-video dataset. Metrics are reported as accuracy. Overall accuracy is computed across all evaluated fields.}
\label{tab:llm_full_dataset}
\small
\setlength{\tabcolsep}{4pt}
\begin{tabular}{lcccccc}
\toprule
\textbf{Model} & \textbf{Overall} & \textbf{Humor Presence} & \textbf{Humor Type} & \textbf{Stand Up} & \textbf{Target Category} \\
\midrule
Gemini 3 Flash Preview   & 0.608 & 0.576 & 0.500 & 0.850 & 0.793 \\
GPT-5 Mini               & 0.591 & 0.542 & 0.469 & 0.744 & 0.807 \\
Claude Sonnet 4.6        & 0.584 & 0.501 & 0.458 & 0.820 & 0.766 \\
Gemini 2.5 Flash         & 0.573 & 0.534 & 0.458 & 0.732 & 0.746 \\
DeepSeek V3.1            & 0.569 & 0.564 & 0.502 & 0.681 & 0.778 \\
\bottomrule
\end{tabular}
\end{table}

\vspace{-0.6cm}
\begin{wrapfigure}{r}{0.48\textwidth}
    \centering
    \vspace{-8pt}
    \includegraphics[width=\linewidth]{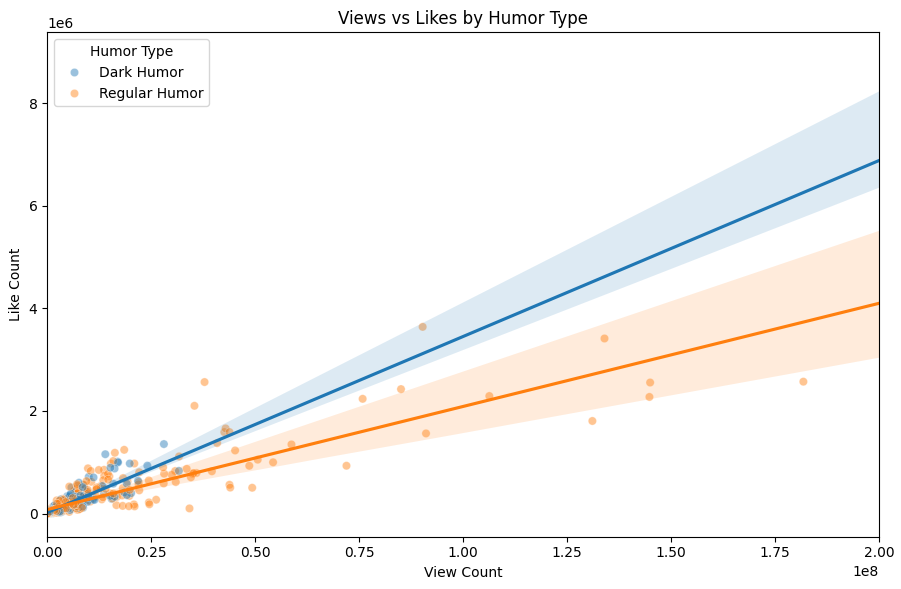}
    \vspace{-8pt}
    \caption{Relationship between view count and like count across humor types. Dark humor content tends to receive more likes for a given number of views, indicating higher audience engagement despite less positive sentiment observed in comments.}
    \label{fig:sentiment_comparison}
    \vspace{-10pt}
\end{wrapfigure}

\subsection{Audience Sentiment by Humor Type}
To better understand how viewers respond to humorous content, we analyzed the sentiment of comments associated with each video using \texttt{twitter-roberta-base-sentiment} model \cite{barbieri2020tweeteval}. We selected this model because it was developed and evaluated for sentiment classification on short, informal social media text, making it well suited for YouTube comments. Its transformer based architecture also allows it to capture contextual cues better than lexicon based methods, which is important when analyzing humor related responses that may include slang, exaggeration, or ambiguity. Comment sentiment was calculated from the top comments associated with each video, meaning the analysis focused on the most visible and relevant audience responses. Regular humor received more positive audience sentiment overall than dark humor. Although the lower quartile of comment sentiment was nearly identical for both categories $-0.150$ for regular humor and $-0.149$ for dark humor, regular humor showed a higher median $0.024$ vs.\ $-0.005$, a higher mean $0.038$ vs.\ $-0.016$, and a higher upper quartile $0.214$ vs.\ $0.120$. These results indicate a trend in which dark humor is associated with more neutral or slightly negative audience reactions, whereas regular humor appears to receive more consistently positive sentiment. 

Interestingly, this pattern differs from engagement signals such as likes (Figure~\ref{fig:sentiment_comparison}), where dark humor content often receives more interaction despite less positive sentiment.

\subsection{Toxicity Patterns in Comment Responses}
From the entire dataset of 1,211 videos, 1,139 videos had available comments, which were used to compute toxicity scores. For each video, comment-level toxicity scores were aggregated to obtain a video-level measure of toxicity. The mean video-level comment toxicity was 0.122. Toxicity values were calculated using the \texttt{detoxify} model and ranged from 0.0009 to 0.948, with a median of 0.110. Detoxify was used because it is a well established model for detecting toxic language in online text. It produces continuous toxicity scores, which makes it appropriate for measuring differences in the severity of YouTube comments across videos rather than reducing responses to a simple toxic versus non toxic label. Its prior use in social media toxicity research also makes it a practical and credible choice for this dataset \cite{hanu2020detoxify}. Comment toxicity also differed by humor type. Dark humor videos had a higher mean comment toxicity than regular humor videos. The values show that dark humor has higher median toxicity comment value of 0.137 compared to regular humor at 0.105. Dark humor also showed greater variability in toxicity with higher standard deviation of 0.092 compared to regular humor. Overall, these results suggest that dark humor tends to attract more toxic comment environments than regular humor. This may be because dark humor often engages with sensitive or controversial themes, which can provoke stronger reactions from viewers, including disagreement, discomfort, or offense, leading to more toxic comment exchanges.

\subsection{Emotion Patterns in Comment Responses}

\begin{wrapfigure}{r}{0.45\textwidth}
    \centering
    \vspace{-5pt}
    \includegraphics[width=\linewidth]{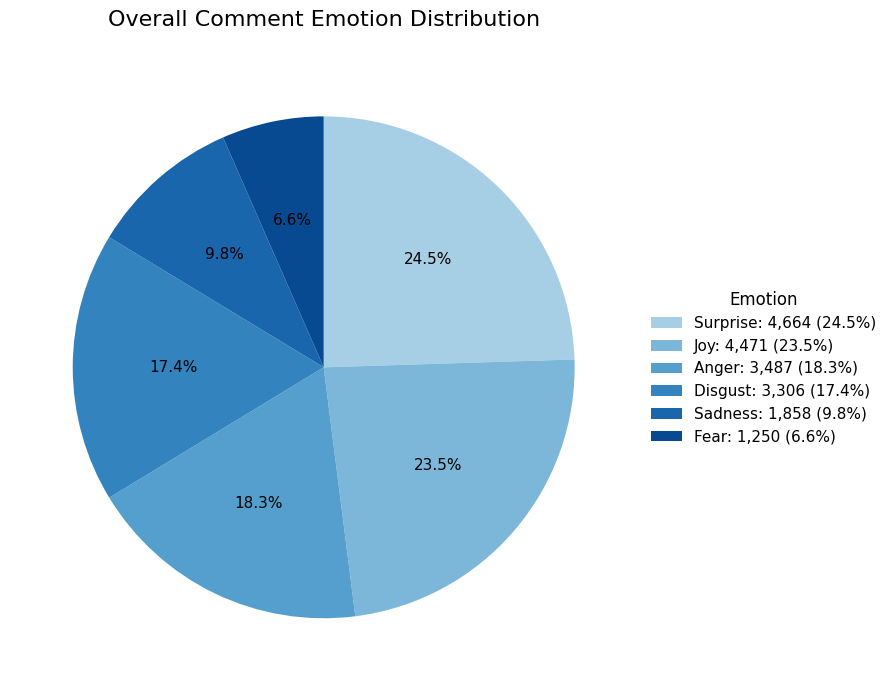}
    \vspace{-5pt}
    \caption{Overall distribution of predicted comment emotions in TwistedHumor. Surprise is the most common emotion, followed closely by joy, with anger and disgust.}
    \label{fig:emotion_distribution}
    \vspace{-5pt}
\end{wrapfigure}

Emotion analysis of the comment corpus further illustrates the ambiguity of audience response to short form humorous content. Figure~\ref{fig:emotion_distribution} summarizes the overall distribution of predicted comment emotions in TwistedHumor and shows that audience responses are not purely positive or negative, but instead span a range of emotions, including \emph{surprise}, \emph{joy}, \emph{anger}, \emph{disgust}, \emph{sadness}, and \emph{fear}.

Neutral emotion was the most common label based on the \texttt{emotion-english-distilroberta-base} model, accounting for 22,418 comments with a mean confidence of 0.711 \cite{hartmann2022emotiondistilroberta}. We used this model because it is specifically designed for English emotion classification and is well suited for short, informal text such as YouTube comments. Unlike sentiment models that mainly capture positive, negative, or neutral polarity, this model provides finer-grained emotion categories such as \emph{joy}, \emph{anger}, \emph{fear}, \emph{sadness}, \emph{surprise}, and \emph{disgust}, which allowed us to better characterize the range of audience reactions to humorous content. 

When comparing across humor types, we notice a clear difference. Among non-neutral emotions, surprise was the most common with 4,664 comments, followed by joy with 4,471 comments. Less common categories were anger, disgust, sadness, and fear. Among regular humor videos, joy, surprise, and sadness were the most common emotions. In contrast, dark humor comments more often reflected disgust, anger, surprise, and fear. Compared with regular humor, dark humor showed lower levels of joy and higher levels of disgust, surprise, and anger. Most notably, the share of anger in dark humor comments was roughly double that of regular humor. This pattern suggests that dark humor is not simply interpreted as another form of comedy, but instead provokes a more conflicted and negatively charged reaction from viewers. This highlights the need to better understand how different audiences interpret such content, especially in settings where context and intent are not explicitly conveyed.

\begin{table*}[t]
\centering
\small
\renewcommand{\arraystretch}{1.2}
\setlength{\tabcolsep}{6pt}
\begin{tabularx}{\textwidth}{p{3.0cm}p{3.0cm}X}
\toprule
\textbf{Analysis focus} & \textbf{LLooM concept from video descriptions} & \textbf{Description} \\
\midrule

\multirow{4}{*}{\textbf{Regular humor category}}
& Humor in Challenges
& Speakers use humor to cope with, frame, or make sense of difficult situations. \\
\cmidrule(l){2-3}

& Critique of Authority
& Humor is directed toward authority figures, institutions, or broader social expectations. \\
\cmidrule(l){2-3}

& Vulnerability in Connection
& Humor reflects emotional struggle, relational tension, and attempts to connect with others. \\
\cmidrule(l){2-3}

& Cultural Social Dynamics
& Humor is shaped by cultural identity and patterns of social interaction. \\

\midrule

\multirow{2}{*}{\textbf{Dark humor category}}
& LLooM vs.\ human annotation
& Based on descriptions alone, LLooM identified 159 videos in the dark humor category, compared to 199 videos annotated as dark humor by human coders. \\
\cmidrule(l){2-3}

& Recurring cluster patterns
& LLooM concept clusters suggest recurring dark humor patterns. Social and emotional framing concepts include \emph{awkward dark humor}, \emph{dark humor in isolation}, and \emph{humor in serious contexts}. Rhetorical style concepts include \emph{ironic dark humor}, \emph{nostalgic dark humor}, and \emph{playful morbid humor}. \\

\bottomrule
\end{tabularx}
\caption{High-level LLooM concepts identified from video descriptions. The analysis reveals broader humor patterns, the most frequent concepts, and recurring dark humor cluster structures.}
\label{tab:lloom_video_descriptions}
\vspace{-1cm}
\end{table*}

\subsection{Topic and Concept Analysis from Descriptions}

To better understand the thematic diversity of the dataset, we performed topic modeling on the video descriptions. 
\begin{wrapfigure}{r}{0.425\textwidth}
    \centering
    \vspace{-5pt}
    \includegraphics[width=\linewidth]{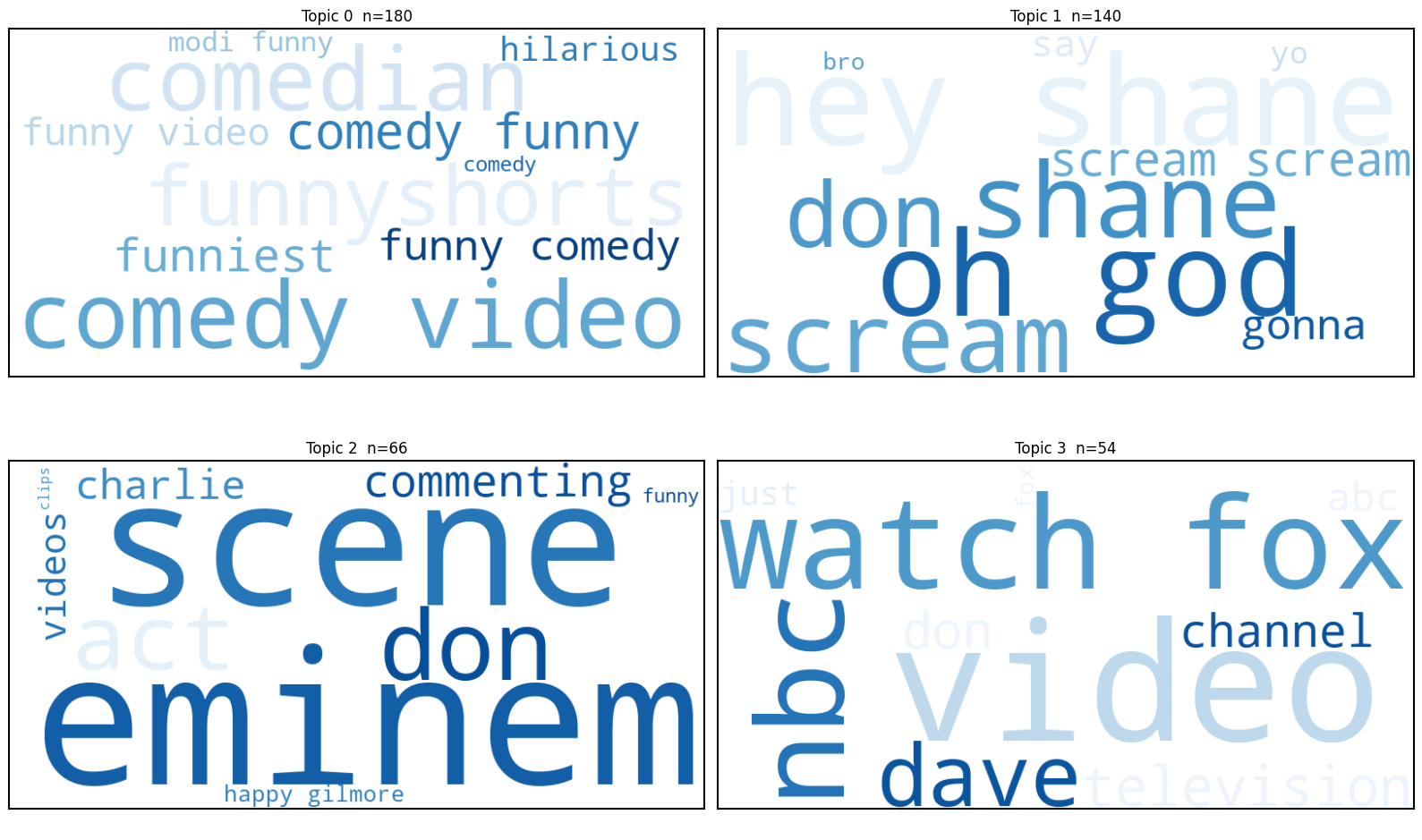}
    \vspace{-5pt}
    \caption{Topic distribution derived from video descriptions. The largest clusters correspond to general comedy content and conversational humor, with additional topics spanning entertainment, television commentary, food, and political content. This highlights the thematic diversity of the dataset.}
    \label{fig:output2}
    \vspace{-5pt}
\end{wrapfigure}
The largest topic cluster (180 videos) was centered on general comedy content, comedians, and funny short videos. The next largest cluster (140 videos) captured conversational and reaction-based humor, while the last topic cluster (66 videos) reflected entertainment. Several additional topics mentioned television commentary, food, and political content (e.g., references to Donald Trump and family). Figure~\ref{fig:output2} shows word-clouds of all topic clusters. Overall, the description-based topic distribution shows that the dataset is not dominated by a single topic and consists of a broad mix of stand-up clips. 

We conducted an LLooM based concept induction analysis on all the video descriptions and transcriptions. LLooM is a concept induction algorithm that leverages large language models to iteratively synthesize sampled text and propose human interpretable concepts of increasing generality \cite{LlooM}. In contrast to topic modeling, which groups content by shared themes, LLooM helps surface higher level conceptual patterns \cite{LlooM}. Table~\ref{tab:lloom_video_descriptions} summarizes the main concepts identified by LLooM. 

Concept induction on video transcripts revealed several higher-level patterns in how humor is expressed in spoken content. These included \textbf{\emph{Humor in Challenges}}, where speakers use humor to cope with or frame difficult situations, and \textbf{\emph{Critique of Authority}}, where humor is directed toward authority figures or social expectations. LLooM also identified \textbf{\emph{Vulnerability in Connection}}, which captures emotional struggle and relational tension, and \textbf{\emph{Cultural Social Dynamics}}, which reflects humor shaped by cultural identity and social interaction. These concepts suggest that transcript-level humor extends beyond surface topics and often functions as a social and emotional mechanism for critique, coping, and identity expression.

In contrast, concept induction on video descriptions highlights recurring thematic patterns. The most common concept identified was humor in serious contexts (48 videos), followed by ironic dark humor (46 videos) and awkward dark humor (36 videos). Based on descriptions alone, LLooM identified 159 videos in the dark humor category, compared to 199 videos annotated as dark humor by human annotators. This difference may be due to the fact that descriptions provide limited context and often omit tone, delivery, and multimodal cues, making it harder for LLooM to detect more implicit forms of dark humor.

Inspection of the LLooM concept clusters shows that dark humor has recurring patterns. Some concepts reflect social and emotional framing, such as awkward dark humor, dark humor in isolation, and humor in serious contexts. Others reflect rhetorical style, such as ironic dark humor, nostalgic dark humor, and playful morbid humor.

\section{Conclusion}
\label{sec:conclusion}

\textbf{Limitations and Future Directions:} This study has several limitations, but also opens clear directions for future work. First, our analysis is limited to YouTube Shorts, and the findings may not generalize to other platforms. Second, although the dataset includes video, transcripts, metadata, and comments, our model evaluation is restricted to transcripts, which do not capture important multimodal cues such as timing, tone, and visual delivery that are central to humor interpretation. Third, annotation of humor and dark humor remains somewhat subjective, and certain boundary cases were challenging even for human annotators. 

At the same time, these limitations highlight the value of the dataset for future research. \corpus\ can support the development of multimodal models that jointly analyze visual, textual, and contextual signals for humor understanding and audience response analysis. Future work can also explore how audience reactions vary across creators, topics, and communities, and investigate how systems can better account for ambiguous or context-dependent content. More broadly, this dataset enables the study of difficult boundary cases, providing a foundation for designing more robust and context-aware approaches to social media safety.

Future work can also investigate how engagement signals such as likes interact with sentiment and toxicity to better understand audience behavior in response to ambiguous humor.

\textbf{Discussion and Final Takeaways:} This study provides three main insights aligned with our research questions. First, we find that dark humor differs structurally from regular humor, with distinct patterns in topics, target categories, and rhetorical devices. Second, our evaluation shows that existing LLMs struggle to reliably identify humor presence and distinguish between regular and dark humor using transcripts alone, highlighting limitations of text-only understanding. Third, audience response analysis reveals that dark humor is associated with more mixed or negative sentiment, higher comment toxicity, and stronger patterns of \emph{anger}, \emph{disgust}, and \emph{fear}, compared with regular humor.

Interestingly, we observe a disconnect between audience sentiment and engagement: dark humor often receives higher interaction (e.g., likes) despite less positive sentiment. This suggests that engagement signals do not necessarily reflect positive reception, and that dark humor may provoke stronger or more polarized reactions that drive interaction.

\clearpage
\bibliographystyle{ACM-Reference-Format}
\bibliography{references}

\end{document}